\documentclass[twocolumn,showpacs,preprintnumbers,amsmath,amssymb]{revtex4}

\usepackage{graphicx}
\usepackage{dcolumn}
\usepackage{bm}

\begin{document}

\preprint{MIT/MKI-07-19}

\title{Neutrino Backgrounds to Dark Matter Searches}

\author{Jocelyn Monroe}
 \email{jmonroe@mit.edu}
\affiliation{%
Department of Physics, Massachusetts Institute of Technology
}%
\author{Peter Fisher}%
\affiliation{%
Laboratory for Nuclear Science, MIT Kavli Institute for Astrophysics and Space Research, Department of Physics, Massachusetts Institute of Technology
}%

\date{\today}

\begin{abstract}
Neutrino coherent scattering cross sections can be as large as
$10^{-39}$ cm$^2$, while current dark matter experiments have
sensitivities to WIMP coherent scattering cross sections five orders
of magnitude smaller; future experiments plan to have sensitivities to
cross sections as small as $10^{-48}$ cm$^2$. With large target masses
and few keV recoil energy detection thresholds, neutral current
coherent scattering of solar neutrinos becomes an irreducible
background in dark matter searches.  In the current zero-background
analysis paradigm, neutrino coherent scattering will limit the
achievable sensitivity to dark matter scattering cross sections, at
the level of $10^{-46}$ cm$^2$.
\end{abstract}

\maketitle

\section{\label{sec:introduction}Introduction}
Dark matter comprises approximately 25\% of the mass of the
universe~\cite{spergel2003}, yet its particle properties are unknown.
Dark matter is observed to interact gravitationally, from which its
density is inferred to be between 0.4 and 0.7 (GeV/c$^2$) per cm$^3$,
and its average velocity $<$$v$$>$ $\simeq$ 230 km/s~\cite{lewin1995}.
The leading dark matter particle candidate is the lightest predicted
supersymmetric particle, the LSP.  Supersymmetry predicts that the LSP
interacts weakly with atomic nuclei, that the LSP mass is in the range
of 10 to 10$^4$ GeV/c$^2$, and that the cross section lies in the
range of $10^{-42}$ to $10^{-48}$ cm$^2$~\cite{ellis2005}.  Collider
experiments have mostly excluded masses below 80 GeV/c$^2$ and cross
sections larger than $\sim 10^{-42}$ cm$^2$ in minimal supersymmetric
models~\cite{ellis2005}.

Direct detection experiments search for dark matter particles using the
coherent elastic scattering process.  Ordinary neutrinos of energy
around 10 MeV also interact coherently with atomic nuclei, causing the
nucleus to recoil with energies up to tens of keV.  Such recoils would
be indistinguishable from dark matter interactions.  The scale of the 
ambient neutrino flux in this energy range is $10^6$ per cm$^{2}$
per second, and the coherent neutrino-nucleus cross section is of
order 10$^{-39}$ cm$^2$.  

In this paper, we estimate background rates in dark matter detectors
caused by coherent neutrino-nucleus elastic scattering of ambient
neutrinos.  We find that $\nu$-$A$ coherent scattering produces 10-100
background events in experiments with few keV thresholds and ton-year
exposures.  In the prevalent zero-background analysis method, this
translates into a fundamental lower bound of roughly 10$^{-46}$ cm$^2$
on the dark matter cross section sensitivity achievable by direct
detection experiments.

\section{\label{sec:detectors}Dark Matter Detection}
Many experiments seek to detect dark matter particles via their
elastic scattering interactions with detector
nuclei~\cite{gaitskell2004}.  The current experimental method is to
set upper limits on the dark matter scattering cross section based on
the observation of zero signal events, using the Yellin gap
technique~\cite{yellin2002}.  This statistical procedure optimizes the
upper limit an experiment can set, at a given confidence level, by
finding the largest possible region of parameter space which contains
zero background events.  In a statistically unbiased way, this method
effectively turns any experiment with low background rates into a
``zero-background'' experiment, albeit over a restricted region of the
experiment's acceptance.  The more background events an experiment
has, the smaller the ``gap'' between events, and the worse the
sensitivity.  Using this approach, recent
observations~\cite{xenon2007,cdms2005} limit the magnitude of the
scattering cross section to be less than approximately $10^{-43}$
cm$^2$.  This corresponds roughly to one background event per kilogram of
detector fiducial mass per day of detector live time.

Dark matter experiments search for a very rare signal process which is
detected via the observation of recoiling nuclei with kinetic energies
as low as 2 keV~\cite{xenon2007}.  Current experiments with masses of
10 kg have zero-background cross section sensitivities of $10^{-44}$
cm$^2$, and project that spanning the predicted range of LSP
interaction cross sections requires 1 ton target masses.  The expected
kinetic energy distribution of recoiling nuclei is exponential,
falling from zero to about 100 keV.  Direct dark matter detection
experiments gain significantly in sensitivity with lower recoil energy
detection thresholds.

Neutrino-nucleus coherent scattering can also produce nuclear recoils
with kinetic energies of a few keV, and, though coherent $\nu$-$A$
scattering has never been observed, the process is theoretically well
understood.  The calculated Standard Model cross section is relatively
large, of order $10^{-39}$ cm$^2$~\cite{freedman1977,drukier1984}.
There has been interest in using this process to make precision weak
interaction measurements at the SNS~\cite{scholberg2006}, to search
for supernova neutrinos~\cite{horowitz2003} and to measure neutrinos
produced in the sun~\cite{cabrera1984}.  Even before direct dark
matter detection experiments existed this process was anticipated as a
background~\cite{drukier1986}.  Here we calculate the background rates
caused by $\nu$-$A$ coherent scattering in target materials relevant
to current dark matter searches.  We consider the recently-measured
solar and geo-neutrino fluxes, and include the nuclear form factors in
the coherent cross section calculation.

\section{\label{sec:bgnd}Neutrino Backgrounds}
Neutrino interactions are an irreducible source of background since no
detector can be shielded from the ambient flux of incident neutrinos.
Direct dark matter detection experiments have pushed the energy
threshold frontier to a few keV, and achieved background levels of
approximately 1 event/kg/keV/day.  In such an environment, $\nu$-$A$
coherent scattering becomes observable, and a source of background to
direct detection dark matter searches.

\subsection{\label{sec:flux}Neutrino Fluxes}
\begin{figure*}
\vspace{-0.5cm}
\hspace{-0.5cm}
\includegraphics[height=7.cm]{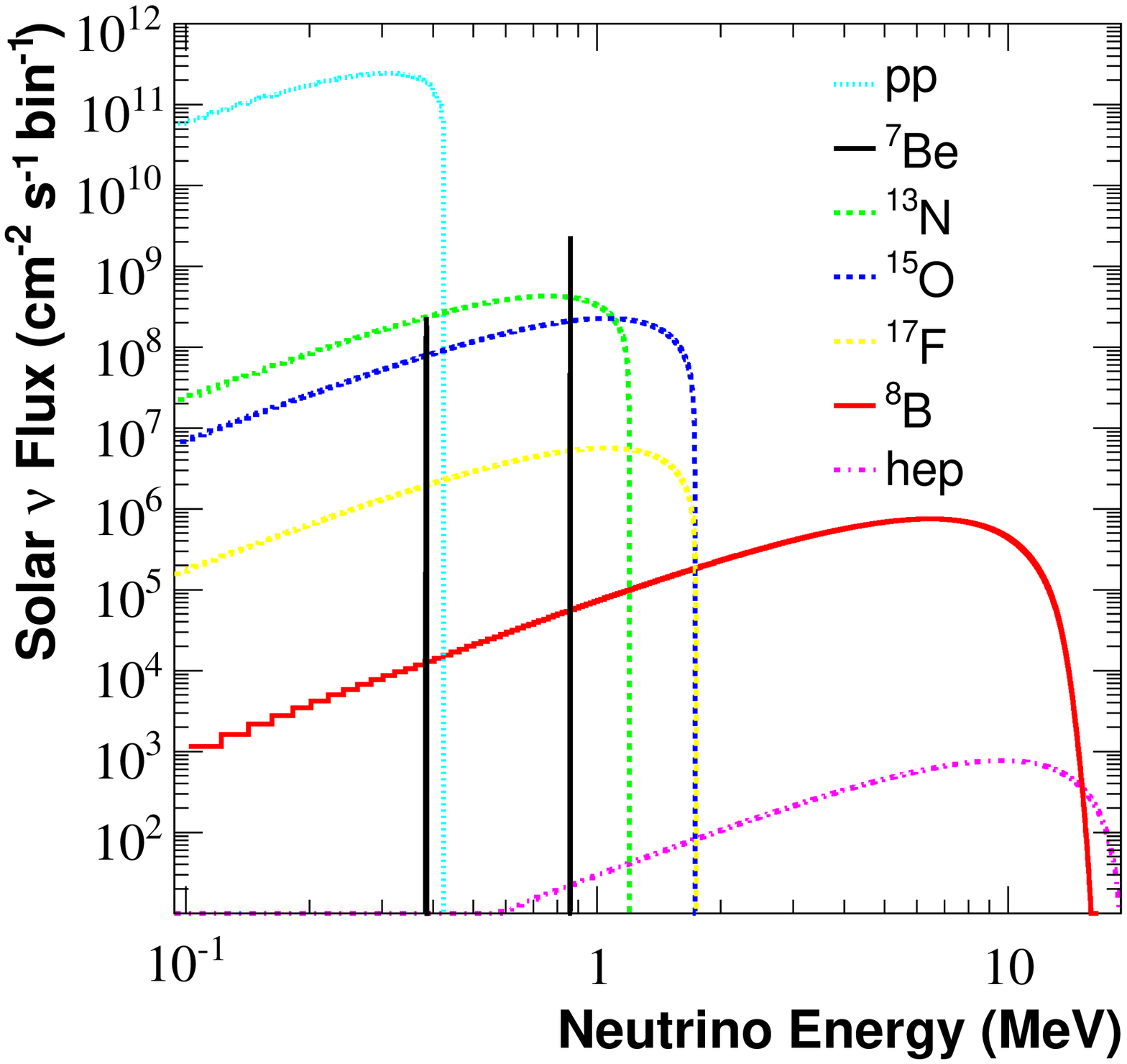}
\hspace{1.cm}
\includegraphics[height=7.cm]{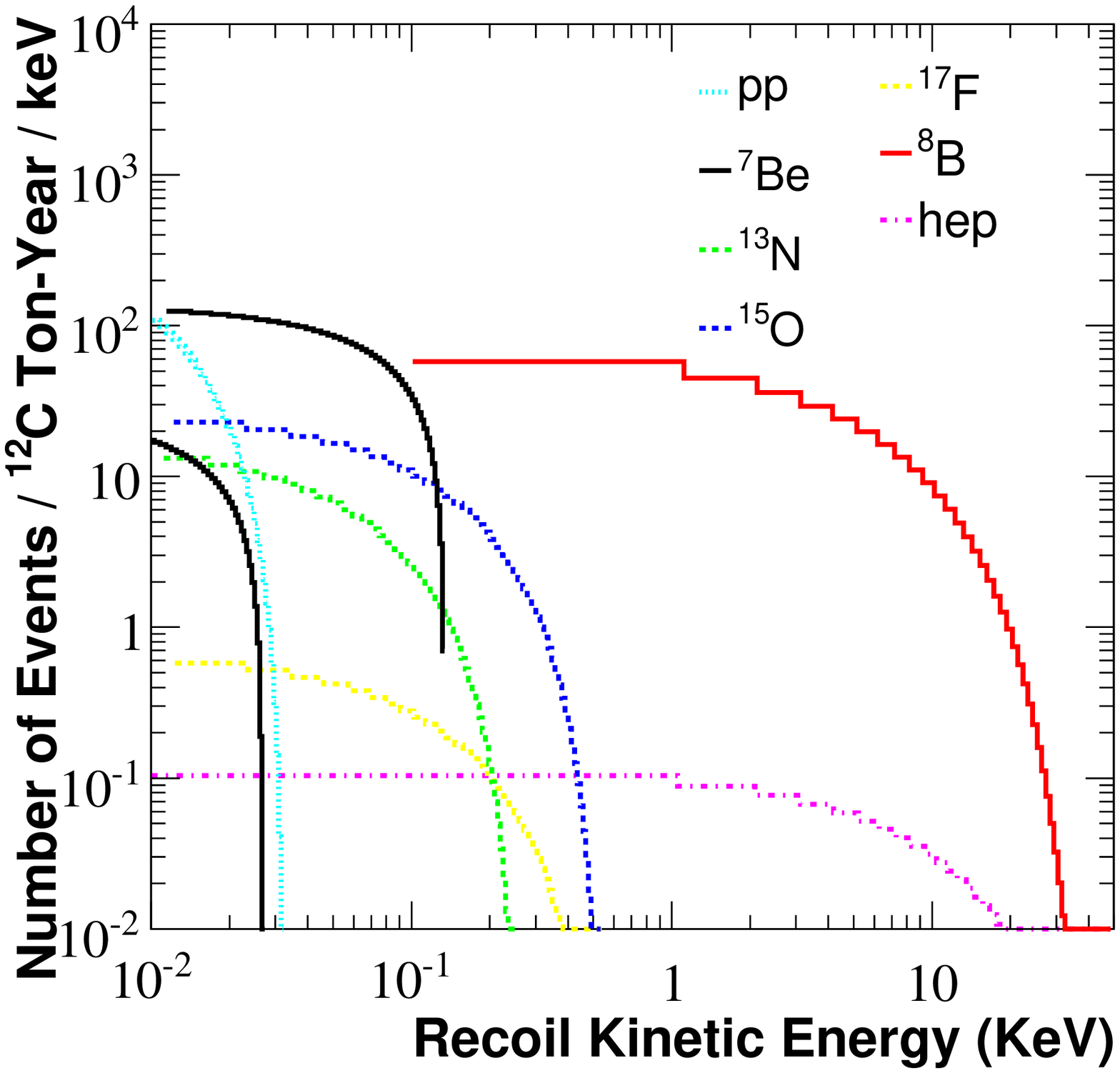}
\vspace{-0.5cm}
\caption{\label{fig:solar} Left: number of solar neutrinos per cm$^2$ per second per energy bin vs. neutrino energy (MeV).  Right: number of solar neutrino-nucleus coherent scattering events per $^{12}$C ton-year exposure, normalized per keV, vs. recoil kinetic energy (keV).}
\end{figure*}

\begin{figure*}
\vspace{-0.5cm}
\hspace{-0.5cm}
\includegraphics[height=7.cm]{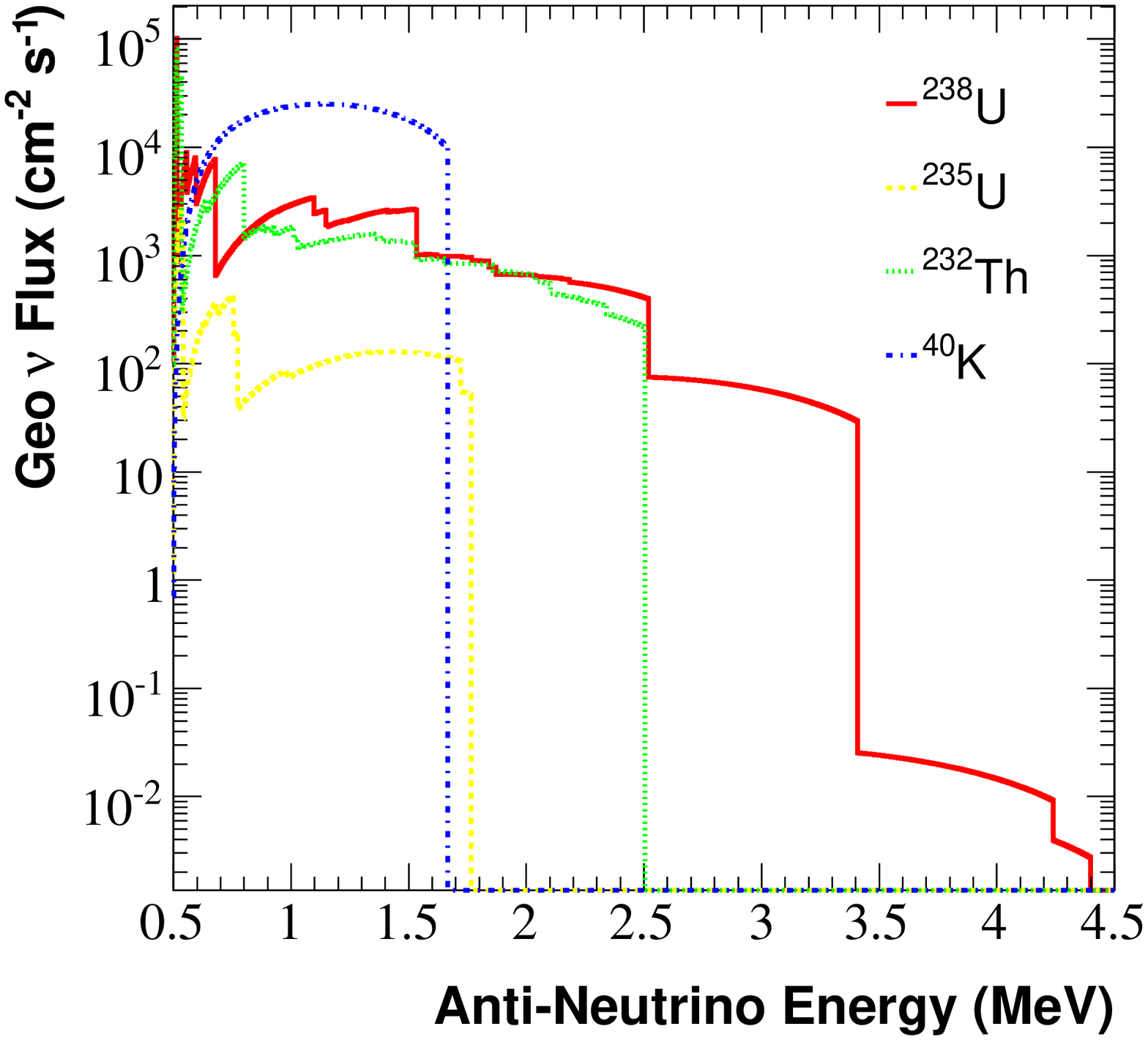}
\hspace{1.cm}
\includegraphics[height=7.cm]{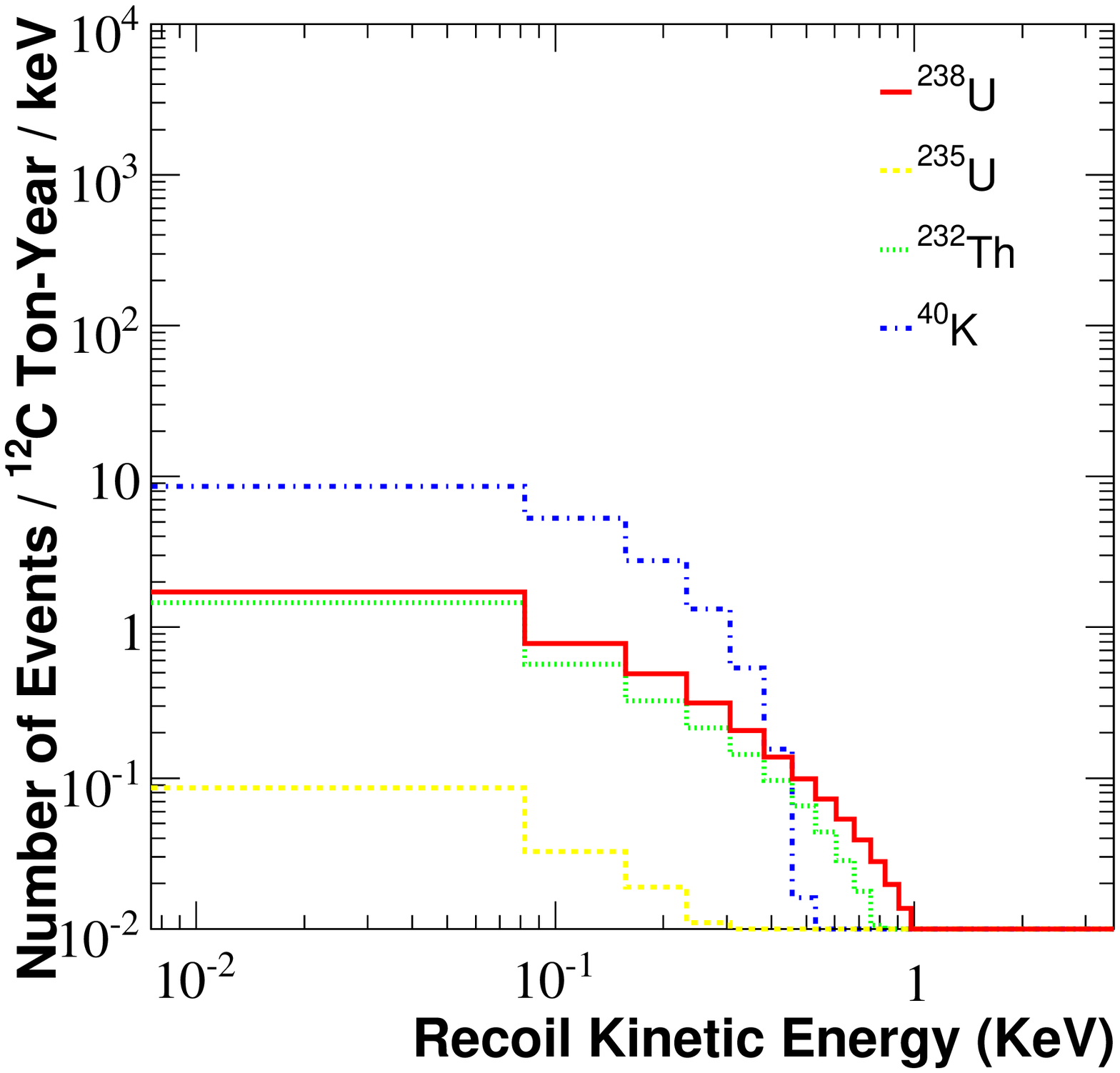}
\vspace{-0.5cm}
\caption{\label{fig:geo} Left: number of geo anti-neutrinos per cm$^2$ per second vs. neutrino energy (MeV).  Right: number of geo neutrino-nucleus coherent scattering events per $^{12}$C ton-year exposure, normalized per keV, vs. recoil kinetic energy (keV).}
\end{figure*}

\begin{figure*}
\vspace{-0.5cm}
\hspace{-0.5cm}
\includegraphics[height=7.cm]{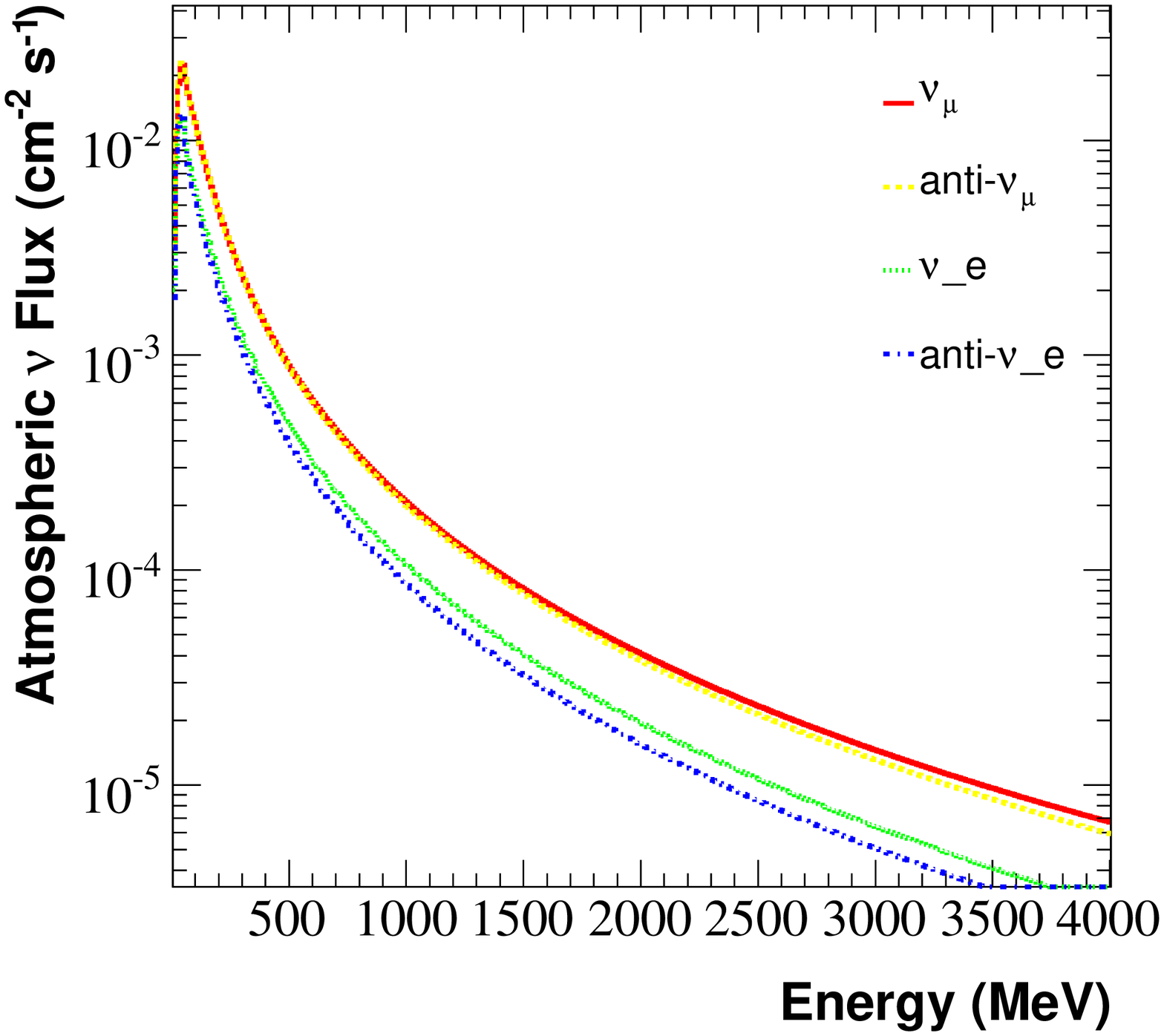}
\hspace{1.cm}
\includegraphics[height=7.cm]{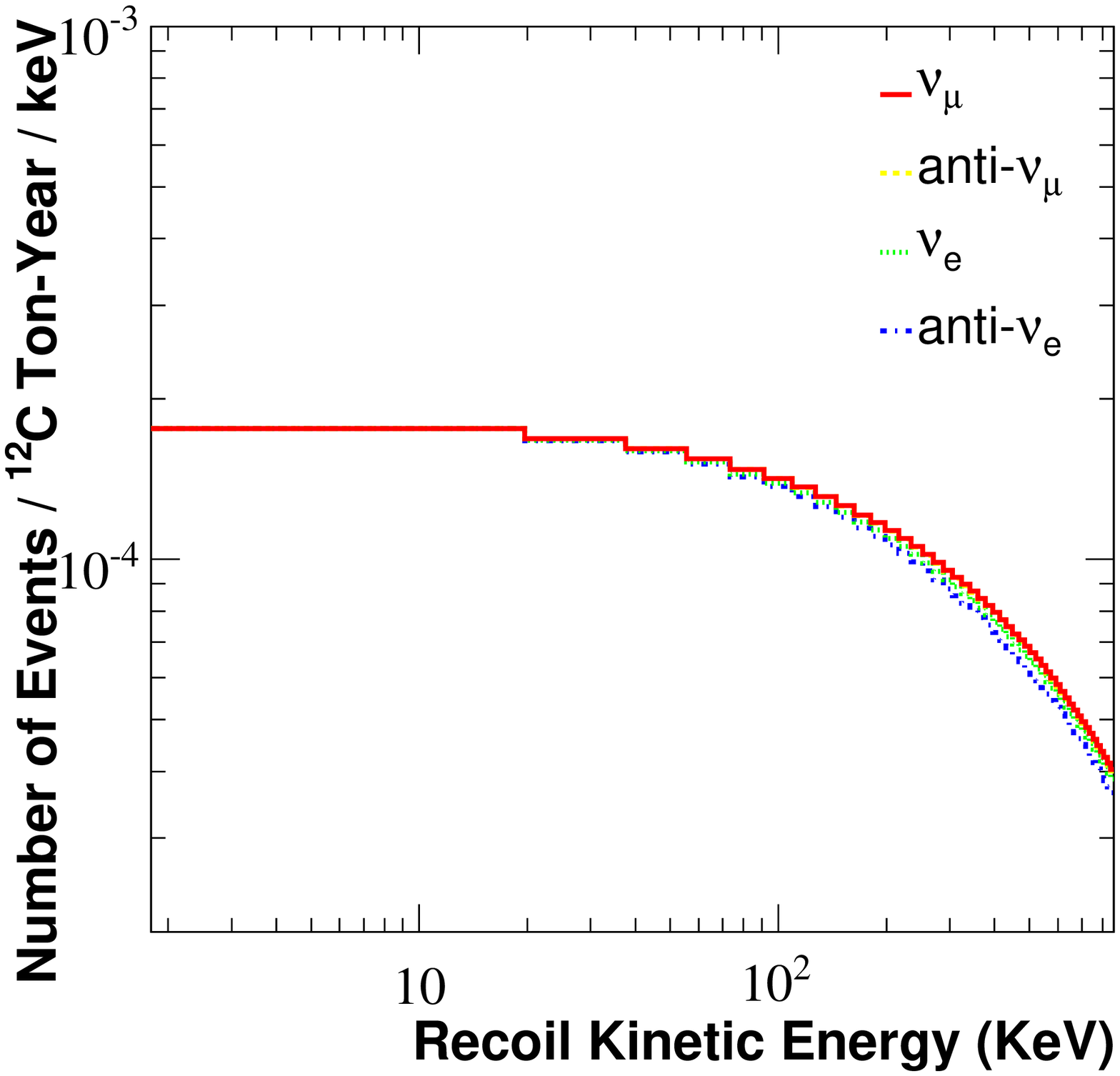}
\vspace{-0.5cm}
\caption{\label{fig:atmo} Left: number of atmospheric neutrinos per cm$^2$ per second vs. neutrino energy (MeV).  Right: number of atmospheric neutrino-nucleus coherent scattering events per $^{12}$C ton-year exposure, normalized per keV, vs. recoil kinetic energy (keV).}
\end{figure*}

The flux of ambient neutrinos and anti-neutrinos is large, with
contributions from many sources.  These include neutrinos produced in
fusion reactions in the sun, anti-neutrinos produced in radioactive
decays in the earth's mantle and core, atmospheric neutrinos and
anti-neutrinos produced by the decays of cosmic ray collision products
in the upper atmosphere, supernova relic anti-neutrinos, and man-made
anti-neutrinos produced in fission processes at nuclear reactors.  The
fluxes of ambient sources of neutrinos and their approximate energy
ranges are shown in table \ref{tab:flux_table}.  Of these, we consider
solar, geo-neutrino, and atmospheric fluxes.  The energy distributions
of the fluxes used in our calculation are shown in figures
\ref{fig:solar}, \ref{fig:geo}, and
\ref{fig:atmo}.
\begin{table}
\caption{\label{tab:flux_table}Ambient sources of neutrinos.  Fluxes are given in number per cm$^2$ per second.}
\begin{ruledtabular}
\begin{tabular}{lll}
Source & Predicted Flux & Energy (MeV)\\
\hline
\cite{bahcall2004} Solar $\nu$ pp      & 5.99$\times10^{10}$ & $<$0.4  \\
\cite{bahcall2004} Solar $\nu$ CNO     & 5.46$\times10^{8}$  & $<$2  \\
\cite{bahcall2004} Solar $\nu$ $^7$Be  & 4.84$\times10^{9}$  & 03, 0.8 \\
\cite{bahcall2004} Solar $\nu$ $^8$B   & 5.69$\times10^{6}$  & $<$12 \\
\cite{bahcall2004} Solar $\nu$ h.e.p.  & 7.93$\times10^{3}$  & $<$18 \\
\hline
\cite{enomoto2005} Geo $\overline{\nu}$ $^{238}$U   & 2.34$\times10^{6}$  & $<$5 \\
\cite{enomoto2005}Geo $\overline{\nu}$ $^{232}$Th  & 1.99$\times10^{6}$  & $<$2.5 \\
\cite{enomoto2005}Geo $\overline{\nu}$ $^{235}$U   & $\sim$4$\times10^{3}$  & $<$2 \\
\cite{enomoto2005}Geo $\overline{\nu}$ $^{40}$K    & $\sim$1$\times10^{7}$  & $<$2 \\
\hline
\cite{gaisser2002} Atmospheric $\nu$+$\overline{\nu}$   & O(1/E(GeV)$^{2.7})$)   & 0-10$^3$ \\
\hline
\cite{nakajima2006} Reactor $\overline{\nu}$       & O(1$\times10^{20}/$distance$^2$) & $<$10 \\
\hline
\cite{ando2003} Supernova Relic $\overline{\nu}$ & O(10$^1$)   & $<$60 \\
\end{tabular}
\end{ruledtabular}
\end{table}

The $^8$B solar neutrino flux is well understood.  The measured
normalization of the total $^8$B solar neutrino flux agrees with the
predicted flux, shown in table \ref{tab:flux_table}, to
2\%~\cite{sno2007}.  Although the predicted flux normalization has an
uncertainty of 16\%~\cite{bahcall2004}, the measured flux, including
neutrino oscillations, has an uncertainty of 3.5\%~\cite{sk2005}.  The
geo-neutrino flux is less well known.  The flux from $^{238}$U and
$^{232}$Th decays has been measured to be approximately 4 times the
predicted magnitude shown in table \ref{tab:flux_table}, with a
measurement uncertainty of 76\%~\cite{kamland2005}.  The atmospheric
neutrino flux is measured by a number of experiments to be consistent
with predictions including neutrino oscillations; the estimated
normalization uncertainty is 10\% for neutrino energies below 100
MeV~\cite{superk2005}.  The atmospheric flux extends over a large
energy range, but only the lowest energy neutrinos are interesting
here since the coherent scattering process occurs for energies below
roughly 50 MeV.  The normalization of the low energy component of the
atmospheric neutrino flux depends strongly on latitude because of the
geo-magnetic cutoff; for example, the flux at the SNO experiment is
approximately 50\% larger than at Super-Kamiokande.  We use the
atmospheric neutrino flux prediction at Super-Kamiokande~\cite{honda2001}.

The calculations here use the predicted solar, geo, and atmospheric
neutrino fluxes, without including neutrino oscillations.  The
coherent scattering process is neutrino-flavor independent to leading
order, and we assume no sterile neutrino participation in
oscillations, thus the oscillated and un-oscillated predicted neutrino
fluxes are in practice equivalent for our calculation.

\subsection{\label{sec:xsec}Neutrino Scattering Cross Sections}

Dark matter experiments are potentially sensitive to two kinds of
neutrino interactions: $\nu$-$e^-$ neutral current elastic scattering,
where the neutrino interacts with the atomic electrons, and $\nu$-$A$
neutral current coherent elastic scattering, where the neutrino
interacts with the target nucleus.  The former process has been
considered as a method for solar neutrino detection in low-threshold
detectors~\cite{bahcall1995}.  The maximum recoil electron kinetic
energy can be as large as a few hundred keV, and the cross sections
are of order $10^{-44}$ cm$^2$.  The latter process has never been
observed since the maximum nuclear recoil kinetic energy is only a few
tens of keV, however, the cross section is relatively large,
approximately $10^{-39}$ cm$^2$.  This work focuses exclusively on
coherent $\nu$-$A$ scattering.

The maximum recoil kinetic energy in $\nu$-$A$ coherent scattering is 
\begin{equation*}
T_{max} \ = \ \frac{2 E_{\nu}^2}{M + 2 E_{\nu}}.
\end{equation*}
where $E_{\nu}$ is the incident neutrino energy, and $M$ is the mass
of the target nucleus.  The four-momentum transfer is related to the
recoil kinetic energy by $Q^2$ = 2$M$$T$, and the three-momentum
transfer $q$ is approximately $\sqrt{2 M T}$.  For neutrino energies
below 20 MeV and nuclear targets from $^{12}$C to $^{132}$Xe, the
maximum recoil kinetic energy ranges from approximately 50 down to 2
keV, and therefore the maximum possible $q$ is quite small, $<$1
fm$^{-1}$.  Typical nuclear radii, $R$, are 3-5 fm, and therefore the
product $q R < 1$.  In this regime, the neutrino scatters coherently
off of the weak charge of the entire nucleus, which is given by
\begin{equation*}
Q_W \ = \ N - (1 - 4 \sin^2 \theta_W) Z
\end{equation*} 
where $N$ and $Z$ are the number of target nucleons and protons
respectively, and $\theta_W$ is the weak mixing angle.  Through the
dependence on $Q_W$, coherence enhances the scattering cross section
with respect to the single nucleon cross section by a factor of $N^2$.

The $\nu$-$A$ coherent scattering cross section is given
by~\cite{freedman1977,drukier1984}
\begin{equation*}
\frac{d\sigma}{d(\cos \theta)} \ = \ \frac{G_F^2}{8\pi} \ Q_W^2 \ E_{\nu}^2 \ (1 + \cos \theta) \ F(Q^2)^2
\end{equation*}
where $G_F$ is the Fermi coupling constant, $Q_W$ is the weak charge
of the target nucleus, $E_{\nu}$ is the projectile neutrino energy,
$\cos \theta$ is the scattering angle in the lab frame of the recoil
nucleus with respect to the incoming neutrino direction, and $F(Q^2)$
is a nuclear form factor that describes the distribution of weak
charge within the nucleus.  In this work, we use form factors
calculated for $^{12}$C, $^{19}$F, $^{40}$Ar, $^{76}$Ge, and
$^{132}$Xe~\cite{horowitz1981}.  The suppression of the cross section
by the nuclear form factor is 5-10\%.

The dependence of the cross section on scattering angle means that
solar neutrino elastic scattering events will, in principle, point
back to the sun.  However, the majority of dark matter detectors do
not have directional sensitivity, and so it is most useful to
calculate event rates as a function of recoil nucleus kinetic energy.
The scattering angle and the recoil kinetic energy are related via
2-body kinematics and the cross section can be expressed in terms of
the kinetic energy, $T$, of the recoiling nucleus as
\begin{equation*}
\frac{d\sigma}{dT} \ = \ \frac{G_F^2}{4\pi} \ Q_W^2 \ M^2 \ (1 - \frac{M T}{2 E_{\nu}^2}) \ F(Q^2)^2.
\end{equation*}

The theoretical uncertainty on the coherent $\nu$-$A$ scattering cross
section comes from nuclear modelling in the form factor calculation;
for neutrino energies of the order of 10 MeV the uncertainty is
expected to be less than 10\%~\cite{horowitz2003}.

\subsection{\label{sec:rates}Background Rates}

\begin{figure*}
\vspace{-0.5cm}
\hspace{-0.5cm}
\includegraphics[height=7.cm]{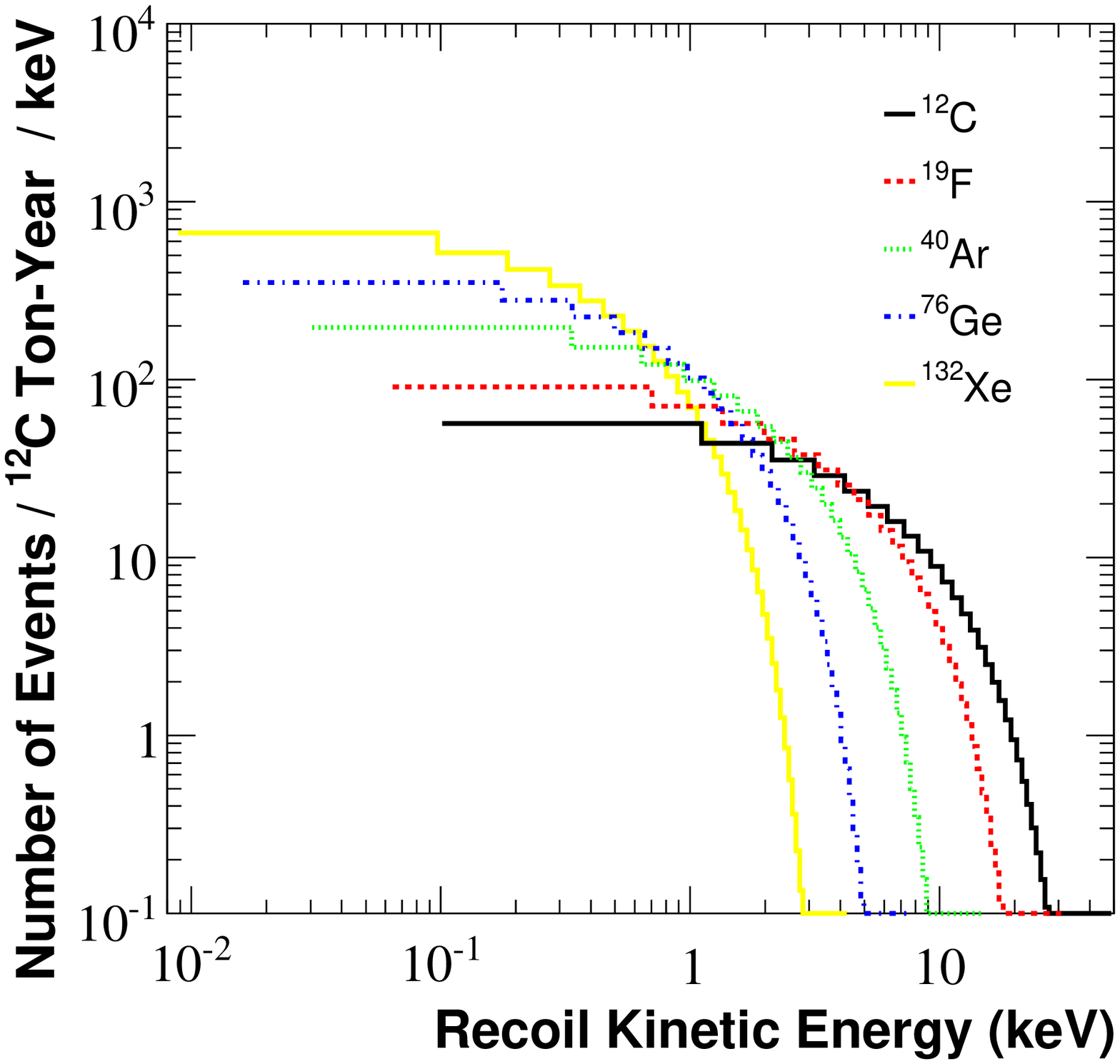}
\hspace{1.0cm}
\includegraphics[height=7.cm]{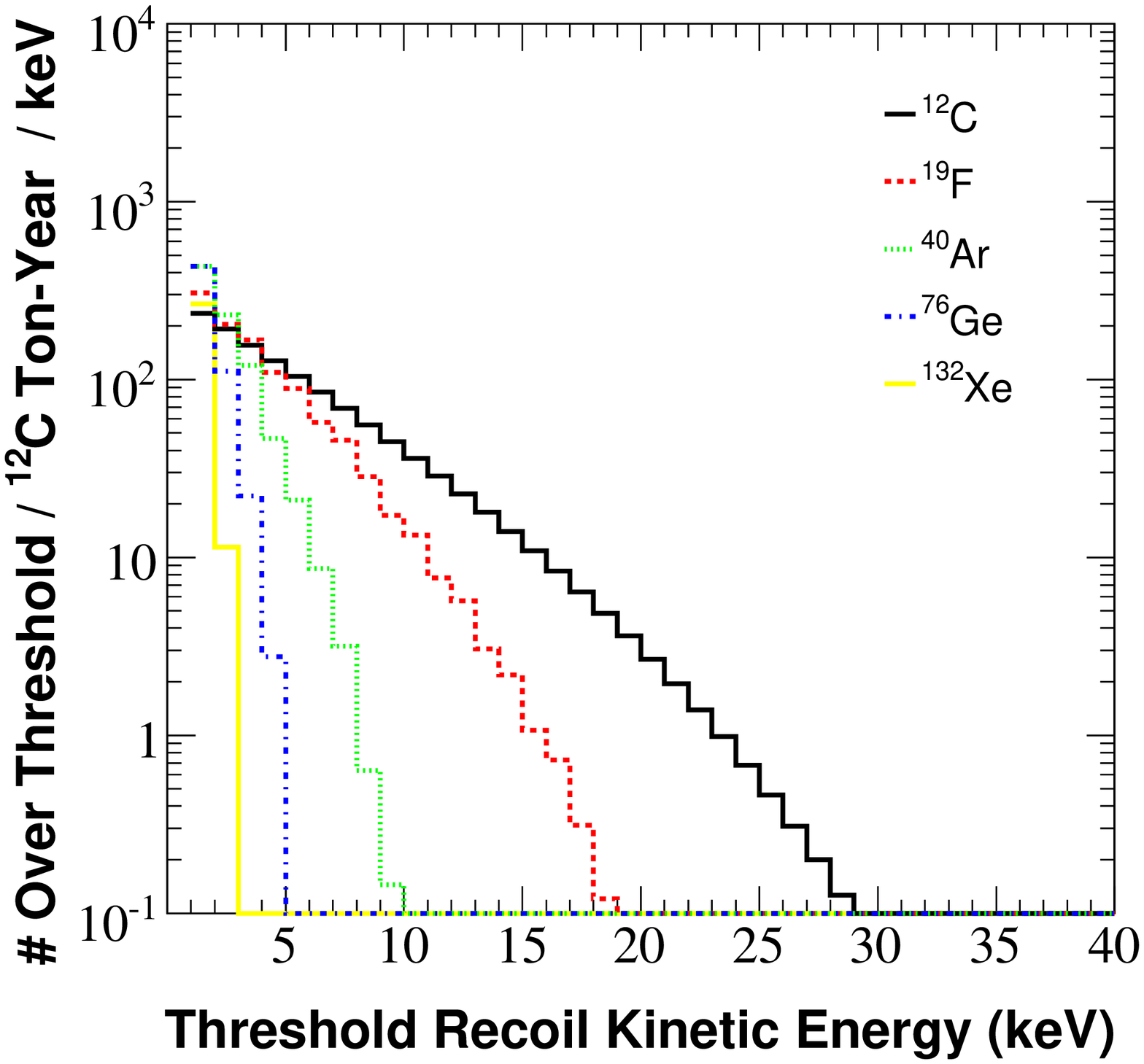}
\vspace{-0.25cm}
\caption{\label{fig:rates_above_threshold} Left: number of $^8B$ solar neutrino-nucleus coherent scattering events per ton-year exposure in various detector media, normalized per keV, vs. recoil kinetic energy (keV).  Right: integrated number of neutrino-nucleus coherent scattering events above threshold per ton-year exposure in various detector media per keV vs. recoil kinetic energy threshold (keV) for $^8B$ solar neutrinos.}
\end{figure*}

With the neutrino fluxes and the $\nu$-$A$ coherent scattering cross
section described above, we calculate the numbers of events per
ton-year exposure as a function of recoil nucleus kinetic energy.
These are shown for a $^{12}$C target in figures
\ref{fig:solar}, \ref{fig:geo}, and \ref{fig:atmo} for solar, geo, and
atmospheric neutrinos respectively.  The recoil energy spectra and
integrated numbers of events over threshold as a function of recoil
energy threshold are compared for $^{12}C$, $^{19}F$, $^{40}Ar$,
$^{76}Ge$ and $^{132}Xe$ in figure \ref{fig:rates_above_threshold}.
\begin{table}
\caption{\label{tab:rates}Rate of $^8$B solar $\nu$-$A$ coherent scattering events per ton-year as a function of minimum nuclear recoil kinetic energy detection threshold.}
\begin{ruledtabular}
\begin{tabular}{lllll}
Target & T$>$0 keV & T$>$2 keV & T$>$5 keV & T$>$10 keV \\
\hline
$^{12}$C   & 235.7  & 191.8 & 104.1  & 36.0   \\
$^{19}$F   & 378.0  & 204.4 & 88.8   & 13.3   \\
$^{40}$Ar  & 804.8  & 231.4 & 21.0   & $<$1.0 \\
$^{76}$Ge  & 1495.0 & 111.5 & $<$1.0 & $<$1.0 \\
$^{132}$Xe & 2616.9 & 14.7  & $<$1.0 & $<$1.0 \\
\end{tabular}
\end{ruledtabular}
\end{table}
The only significant source of events above recoil energies of 1 keV
comes from $^8$B solar neutrinos.  For this source, we summarize the
number of $\nu$-$A$ coherent scattering events per ton-year exposure
for several target nuclei used in current direct detection dark matter
experiments in table \ref{tab:rates}.  For lighter target nuclei,
above a 2 keV threshold, we find that there will be a few hundred
background events to dark matter searches from $\nu$-$A$ coherent
scattering.  This source of background is smaller, for the same
threshold, in heavier target nuclei owing to lower allowed maximum
recoil kinetic energies.

\section{\label{sec:end}Discussion}

For any detector medium, with a ton-year exposure and few keV recoil
energy threshold, solar $\nu$-$A$ coherent scattering will be an
irreducible background to dark matter searches, at the level of 10-100
events depending on the detector energy threshold.  Under the
zero-background assumption, in a counting-only analysis, these events
would be mistaken for a signal.  Following~\cite{lewin1995}, one would
expect 5-25 signal events per ton-year if the cross section were
10$^{-46}$ cm$^2$.  If signal and background cannot be distinguished,
a background of 10-100 $\nu$-$A$ events per ton-year sets a lower
bound on the experimental sensitivity to the true dark matter
scattering cross section.  Thus, there is a fundamental
limit of 10$^{-46}$ cm$^2$ achievable with the zero-background
counting-only method.

This $\nu$-A coherent background could be easily eliminated by
requiring recoil energies greater than the allowed values for coherent
$\nu$-$A$ scattering.  Imposing a cut of $T$$>$30 keV for light
targets, or $T$$>$5 keV for heavier targets, would suffice; however,
this approach would reduce the sensitivity of dark matter searches
approximately as $exp (-\Delta E_{th}/E_0 r)$~\cite{lewin1995}, where
$\Delta E_{th}$ is the change in the recoil energy threshold, $E_0$ is
the kinetic energy of the dark matter particle, and $r$ is $(4 m_D
m_{target})/(m_D + m_{target})^2$.  For example, in a $^{12}C$
detector, increasing the threshold from 5 to 30 keV would reduce the
sensitivity by a factor of 6 for a dark matter particle of mass $m_D$
= 100 GeV/c$^2$.

Abandoning the zero-background paradigm, it may be possible to
discriminate statistically against $\nu$-$A$ coherent scattering
events using the angular distribution, or the recoil kinetic energy
spectrum.  An important caveat is that the sensitivity of experiments
with background events increases with exposure time as $\sqrt{t}$,
whereas without backgrounds the sensitivity is proportional to $t$.  A
standard technique is to search for an excess above a background
expectation; in this case, the uncertainties on the rate and
kinematics of solar $\nu$-$A$ coherent scattering become very
important.  If there is a sizeable signal, a fit to the recoil energy
spectrum could distinguish between the slopes expected from coherent
neutrino scattering and a dark matter signal excess.  Further, a
detector with directional sensitivity and tens of events could fit in
two dimensions: recoil kinetic energy vs. recoil track angle with
respect to the sun.
\begin{acknowledgments}
This work was supported by the MIT Pappalardo Fellowship program and the MIT Kavli Institute.  We wish to thank Kate Scholberg, Chuck Horowitz, and Joe Formaggio for their useful input and helpful discussions.
\end{acknowledgments}

\bibliography{nubgnd}

\end{document}